\documentclass[a4paper,11pt]{article}
\pdfoutput=1 
\usepackage{jcappub} 
\usepackage[T1]{fontenc} 
\usepackage[british]{babel}
\usepackage{xcolor}
\usepackage{booktabs}
\usepackage{latexsym}
\usepackage{amssymb}
\usepackage{amsmath}
\usepackage{graphicx}
\usepackage{tabularx}
\usepackage{cancel}
\usepackage{hyperref}
\usepackage[]{cleveref}
\usepackage[utf8]{inputenc}
\usepackage[normalem]{ulem}
\newcommand{\dd}{\text{d}}

\newcommand{\varpar}{{\mkern3mu\vphantom{\perp}\vrule depth 0pt\mkern2mu\vrule depth 0pt\mkern3mu}}

\title{\boldmath Cosmological coupling of nonsingular black holes}

\author[a,b]{M. Cadoni}
\author[a,b,1]{A. P. Sanna,\note{Corresponding author.}}
\author[a,b]{M. Pitzalis, }
\author[c,d]{B. Banerjee, }
\author[c,d]{R. Murgia, }
\author[c,d]{N. Hazra}
\author[c,d]{and M. Branchesi}

\affiliation[a]{Dipartimento di Fisica, Universit\`a di Cagliari, Cittadella Universitaria, 09042 Monserrato, Italy}
\affiliation[b]{INFN, Sezione di Cagliari, Cittadella Universitaria, 09042 Monserrato, Italy}
\affiliation[c]{Gran Sasso Science Institute (GSSI), Viale F. Crispi 7, L'Aquila (AQ), I-67100, Italy}
\affiliation[d]{INFN - Laboratori Nazionali del Gran Sasso (LNGS), L'Aquila (AQ), I-67100, Italy}

\emailAdd{mariano.cadoni@ca.infn.it}
\emailAdd{asanna@dsf.unica.it}
\emailAdd{mirko.pitzalis@ca.infn.it}
\emailAdd{biswajit.banerjee@gssi.it}
\emailAdd{riccardo.murgia@gssi.it}
\emailAdd{nandini.hazra@gssi.it}
\emailAdd{marica.branchesi@gssi.it}

\abstract{We show that -- in the framework of general relativity (GR) -- if black holes (BHs) are singularity-free objects, they couple to the large-scale cosmological dynamics. We find that the leading contribution to the resulting growth of the BH mass ($M_{\rm BH}$) as a function of the scale factor $a$ stems from the curvature term, yielding $M_{\rm BH} \propto a^k$, with $k=1$. We demonstrate that such a linear scaling is universal for spherically-symmetric objects, and it is the only contribution in the case of regular BHs. For nonsingular horizonless compact objects we instead obtain an additional subleading model-dependent term. We conclude that GR nonsingular BHs/horizonless compact objects, although cosmologically coupled, are unlikely to be the source of dark energy.
We test our prediction with astrophysical data by analysing the redshift dependence of the mass growth of supermassive BHs in a sample of elliptical galaxies at redshift $z=0.8 -0.9$. We also compare our theoretical prediction with higher redshift BH mass measurements obtained with the James Webb Space Telescope (JWST). We find that, while $k=1$ is compatible within  $1 \sigma$ with JWST results, the data from elliptical galaxies at $z=0.8 -0.9$ favour values of $k>1$. New samples of BHs covering larger mass and redshift ranges and more precise BH mass measurements are required to settle the issue.}

\begin{document}
\maketitle
\flushbottom

\section{Introduction}{\label{Introduction}}
Recently, Farrah et al. \cite{Farrah:2023opk}, following previous proposals \cite{McVittie:1933zz,Faraoni:2007es,Croker:2019mup,Croker:2021duf} and building on recent experimental results \cite{Farrah_2023}, found observational evidence for cosmologically-coupled mass growth in supermassive black holes (SMBHs) at the center of elliptical galaxies at different redshift. To do so, they employed the  mass formula (firstly proposed in Ref.~\cite{Croker:2019kje})
 \begin{equation}
     M(a) = M(a_i) \left(\frac{a}{a_i} \right)^k\, , \qquad \text{with} \quad a\geq a_i\, ,
     \label{Farraformula}
 \end{equation}
where $a$ is the scale factor, and $a_i$ is the value of the latter evaluated at a particular reference time. Physically, this is identified as the initial time at which the compact object formed and became coupled to the cosmological background. The slope parameter $k$ is a constant, quantifying the strength of the coupling between the small-scale matter inhomogeneities and the large-scale cosmological dynamics. The physically relevant values of $k$ are limited by causality to the interval $-3 \leq k \leq 3$ \cite{Croker:2021duf}, with $k=0$ corresponding to zero coupling.

The authors of Ref.~\cite{Farrah:2023opk} found a preference consistent with $k \sim 3$ at $90 \, \%$ confidence level (C. L.), pointing therefore to nonzero coupling. This would also imply that the density of these objects in a cosmological volume $a^3$ is constant, thus leading to the conclusion of Ref.~\cite{Farrah:2023opk} that BHs could be responsible for the observed accelerated expansion of the universe (see, however, Refs.~\cite{Rodriguez:2023gaa,Parnovsky:2023wkc,Avelino:2023rac,Mistele:2023fds,Wang:2023aqe,Andrae:2023wge,Lei:2023mke,Amendola:2023ays,Gao:2023keg,Gaur:2023hmk}).

More recently, tests of cosmological BH-mass coupling have been performed using higher redshift Active Galactic Nuclei (AGN) data coming from the JWST survey~\cite{Lei:2023mke}. Strikingly, these tests favour much smaller values of $k$, with a best-fit value of $k\simeq -1$.
Moreover, a $k=3$ cosmological coupling seems to be in tension with the predictions of the LIGO-Virgo-KAGRA network on the rate and typical masses of black hole mergers ~\cite{Ghodla:2023iaz}. In view of this intricate situation, an exhaustive theoretical understanding of the cosmological coupling of BHs and other compact astrophysical objects is clearly crucial.

The main purpose of this letter is to investigate the cosmological coupling of nonsingular compact objects, such as regular BHs and horizonless configurations, in an exact general relativity (GR) framework. This is realized by sourcing Einstein's field equations with an anisotropic fluid (see, \textit{e.g.}, Ref.~\cite{Cadoni:2022chn}, and references therein), which allows to circumvent Penrose's singularity theorem \cite{Penrose:1964wq}.

We start from a parametrization of the spacetime metric that naturally couples the cosmological dynamics with local, spherically-symmetric objects, since BH angular momentum effects are expected to be negligible on cosmological scales. We then show that Einstein's equations allow for solutions describing nonsingular BHs, whose Misner-Sharp (MS) mass is given by the universal form \eqref{Farraformula}, but with $k=1$. In the case of horizonless objects, instead, we find subleading model-dependent corrections to this formula (not considered in Ref. \cite{Farrah:2023opk}). 
We apply our cosmologically-coupled mass expression to some explicit models of regular objects, representing BH mimickers.

In order to test our theoretical prediction with astrophysical data, we perform an analysis of the redshift dependence of the mass growth of SMBHs in elliptical galaxies, along the lines of the study performed in Ref.~\cite{Farrah_2023}. Our sample consists of $108$ SMBHs, at redshift $z= 0.8-0.9$.
We also compare our theoretical result with the outcome of the tests performed using higher redshift AGN data from JWST~\cite{Lei:2023mke}. We find that, while $k=1$ is compatible with the latter within a $1 \sigma$ confidence interval, it is quite in tension with the data from elliptical galaxies, which favour higher values of $k$ instead.

\section{Cosmological coupling of astrophysical compact objects}
The anisotropic fluid we employ to source the gravitational field is described by the stress-energy tensor \footnote{Throughout the entire paper we use natural units $c = \hbar =1$.}
\begin{equation}
T_{\mu\nu} = \left(\rho + p_{\perp} \right)u_{\mu}u_{\nu} + p_{\perp} \ g_{\mu\nu} - \left(p_{\perp}-p_{\varpar} \right)w_{\mu}w_{\nu}\, ,
\label{TensoreEI}
\end{equation}
where $u_{\mu}$ and  $w_{\mu}$ satisfy $g^{\mu\nu} u_{\mu}u_{\nu} = -1$, $g^{\mu\nu} w_{\mu}w_{\nu} = 1$, $u^{\mu}w_{\mu} = 0$. $\rho$, $p_{\varpar}$,  $p_{\perp}$ are, respectively, the fluid density and the pressure components, parallel and perpendicular to the spacelike vector $w_{\mu}$. 

We consider a general parametrization of the spacetime metric, which is suitable to describe the cosmological coupling of local, spherically-symmetric objects:
\begin{equation}
\dd s^2 = a^2(\eta)\left[-e^{\alpha(\eta, r)} \dd t^2 + e^{\beta(\eta, r)} \dd r^2 + r^2 \dd\Omega^2\right]\, .
\label{generalmetric}
\end{equation}
$\alpha(\eta, r)$ and $\beta(\eta, r)$ are metric functions depending on both $\eta$ (the conformal time) and $r$, while $a(\eta)$ is the scale factor.
It was shown in Ref.~\cite{Cadoni:2020jxe} that this parametrization of the metric, together with \cref{TensoreEI}, can be used to describe different regimes: the dynamics of cosmological inhomogeneities on small scales, the dynamics of large cosmological scales, but also their coupling at intermediate scales. 

Appropriate combinations of Einstein's equations give (see Ref.~\cite{Cadoni:2020jxe} for further details) $\dot{\alpha}\dot{\beta} =0$, where the dot means derivation with respect to (w.r.t.) the conformal time. The field equations allow, therefore, for two classes of solutions, characterized either by $\dot{\beta}=0$ or by $\dot{\alpha}=0$. The first case corresponds to a complete decoupling between the cosmological, time-dependent dynamics (that of the scale factor) from the small, $r$-dependent one, pertaining instead to the local inhomogeneity. This model allows for both  standard Friedmann-Lema\^{i}tre-Robertson-Walker (FLRW) and other solutions, but not for black-hole ones in a cosmological background (for further details, see Ref.~\cite{Cadoni:2020izk}).

As we are here interested in a possible cosmological coupling of BHs, we will consider the $\dot{\alpha}=0$ case only. Then, the field equations can be rewritten in the form (see Ref.~\cite{Cadoni:2020jxe}) 

\begin{subequations}
\begin{align}
&e^{-\beta(r,\eta)}= g(r) a^{r\alpha'}\, ; \label{betafunction}\\
&\frac{\dot{a}^2}{a^2}\left(3-r\alpha' \right)e^{-\alpha} + \frac{1-e^{-\beta}+r\beta'e^{-\beta}}{r^2} = 8\pi Ga^2 \rho\, ; \label{alpha00}\\
&\frac{e^{-\beta}+re^{-\beta} 	\alpha'-1}{r^2}+e^{-\alpha}\left(-2\frac{\ddot{a}}{a}+\frac{\dot{a}^2}{a^2} \right)=8\pi G a^2 p_{\varpar}\, ; \label{alpharr}\\
&\dot{\rho} + \frac{\dot{a}}{a}\left(3\rho+3p_{\varpar}+rp'_{\varpar} \right)=0\, , \label{alphaconserv}
\end{align}
\label{systemalphadotzero}
\end{subequations}
where the prime refers to derivation w.r.t. $r$, while $g(r)$ is an integration function. The remaining equation, stemming from the conservation of the stress-energy tensor, is  used to compute $p_\perp$. The system \eqref{systemalphadotzero} has to be closed by imposing additional external conditions like, \textit{e.g.}, the equation of state (EoS) of the fluid. 

The large-scale, cosmological dynamics, determining the scale factor $a$, can be completely decoupled from the equations. In this limit, our universe becomes homogeneous and isotropic, so that the dependence on the radial coordinate $r$ is negligible. Moreover, we will only consider asymptotically flat sections at constant time, so that, for $r\to \infty$, $e^\alpha$, $e^{-\beta}\to 1$. 
In this limit, \cref{generalmetric} describes a spatially flat, FLRW metric, solution of the system \eqref{systemalphadotzero}, which now reduces to the Friedmann equations
\begin{subequations}
\begin{align}
&3\frac{\dot a^2}{a^2} = 8\pi G a^2 \rho_1(\eta) \, ; \label{Fried1}\\
&\frac{\dot a^2}{a^2}-2\frac{\ddot a}{a} = 8\pi G a^2 p_1(\eta)\, , \label{Fried2}
\end{align}
\label{systemFried}
\end{subequations}
where $\rho_1$, $p_1$ are the usual energy density and pressure at cosmological scales, respectively.

Let us now consider the local, small-scale limit of  \cref{systemalphadotzero}, describing compact, spherically-symmetric objects at a time scale for which both $\alpha$ and $\beta$ are almost independent of time.
The system \eqref{systemalphadotzero} describes now a static, spherically-symmetric spacetime, where \cref{betafunction} reads $e^{-\beta(r,\, \eta_0)}\equiv e^{-\beta_0(r)}= g(r) \,  a_\text{R}^{r\alpha'}$, where $\eta_0$ represents a generic reference time. Inverting the above yields the function $g(r)$, \textit{i.e.}, $\ln g = -\beta_0 - r\alpha' \ln a_\text{R}$. 

The constant $a_\text{R}$ gives the scale factor at the reference time, which can be fixed as  the present time, namely $a_\text{R} = 1$. \Cref{betafunction}, thus, becomes
\begin{equation}
    e^{-\beta(r, \eta)} = e^{-\beta_0(r)} a(\eta)^{k(r)}\, ,
    \label{betageneral}
\end{equation}
where we defined
\begin{equation}
k(r) \equiv r \alpha' (r)\, . 
\label{kalpha}
\end{equation}

At fixed time, the system of equations \eqref{systemalphadotzero} reduces to
\begin{subequations}
\label{SystemEqFirstRegime}
\begin{align}
&\frac{1-e^{-\beta}+r\beta'e^{-\beta}}{r^2} = 8\pi G a^2_i \, \rho\, ; \label{alpha00static}\\
&\frac{e^{-\beta}+re^{-\beta} 	\alpha'-1}{r^2}=8\pi G  a^2_i\, p_{\varpar}\, ; \label{alpharrstatic}\\
&\dot{\rho} =0\, , \label{conservstatic}
\end{align}
\end{subequations}
where $a_i$ is the scale factor at a particular, but arbitrary, reference time $\eta_i$. In order to describe the cosmological coupling, $\eta_i$ is  identified as the initial time at which the compact object forms and couples to the cosmological background (see also Croker \emph{et al} \cite{Croker:2021duf,Farrah:2023opk}). 
These equations allow for any static GR solutions, like singular or nonsingular BHs and compact objects (see, \textit{e.g.}, Refs.~\cite{Cardoso:2019rvt,Cadoni:2022chn,Kumar:2021oxa} and references therein). The specific EoS of the fluid sourcing the static solution at fixed time determines the relation between $\beta_0$ and $\alpha$. The explicit form of the latter is, however, not constrained and is determined by the density profile $\rho(r)$. 

Once the cosmological dynamics and the local compact solutions have been determined by \cref{systemFried} and \cref{SystemEqFirstRegime} respectively, \cref{systemalphadotzero} describes a cosmologically-coupled compact object. In fact, \cref{alpha00,alpharr} give now the density $\rho(r,\eta)$ and pressure $p_{\varpar}(r,\eta)$ in terms of $\rho_1(\eta)$, $p_1(\eta)$, $\alpha(r)$, $\beta_0(r)$ and $a(\eta)$. 

We now focus on \cref{alpha00}, which can be written as
\begin{equation}
8\pi G \rho = \frac{8\pi G}{3}\rho_1 \, \frac{1}{r^2}\partial_r\left(r^3 e^{-\alpha} \right) + \frac{1}{a^2r^2} \partial_r \left(r-re^{-\beta} \right) \, .
\label{rhobeforeint}
\end{equation}

Using this equation, we can easily compute the comologically-coupled MS mass of the compact object, contained in a cosmological volume $a(\eta)^3 L^3$, where $L$ is the radius of the compact object, measured in the local $r$ coordinate. We have 

\begin{equation}\begin{split}
M(\eta) &= 4\pi a^3(\eta) \int_0^L \dd r \, r^2 \, \rho(r, \eta) = \frac{4\pi }{3}\rho_1 a^3 L^3 \, e^{-\alpha(L)} + M(a_i) \frac{a}{a_i} \left[1-e^{-\beta_0(L)}a^{k_L} \right]\, ,
\label{MSmasscosmologygeneral}
\end{split}\end{equation}

where $k_L \equiv k(L)$ comes from \cref{kalpha}. Here we defined $M(a_i) \equiv a_i L/2 G$, which is the mass of the object computed at the coupling epoch. Peculiarly, this expression defines the proper Schwarzschild radius $a_i L = 2G M(a_i)$ at the initial time at which the compact object formed and became cosmologically coupled. 

\Cref{MSmasscosmologygeneral} is our most important result. It shows that the mass of compact astrophysical objects is naturally coupled to the cosmological dynamics.  With our formula, we are essentially describing the redshifting of the mass of the object with respect to its formation  epoch.

Let us now briefly discuss the different contributions appearing in Equation~\eqref{MSmasscosmologygeneral}. 

The first term gives a purely cosmological contribution, which depends on the background cosmological energy density $\rho_1$ and is therefore expected to be relevant only beyond the transition scale to homogeneity and isotropy. It will not play a role at the typical scales $L$ of the compact object and, thus, we will not consider it in the reasoning that follows \footnote{This term, depending on the redshift factor $e^{-\alpha}$, gives a divergent contribution if horizons are present, due to our choice of the radial coordinate $r$. Since the MS mass is a covariant quantity \cite{Hayward:1994bu}, any physical result based on its use is independent of the coordinate system chosen. Therefore, the coordinate singularity at the horizon can be removed by choosing  suitable coordinates, like Painleve-Gullstrand coordinates or the $(T, R)$ coordinates adopted in Ref.~\cite{Faraoni:2015uma}, where the MS mass depends only on the $g_{RR}^{-1}$ component and no  coordinate singularities  are present (see also Ref.~\cite{Cadoni:2023lqe} for a recent and thorough discussion on the topic).}.
The second term has the form of a ``universal cosmological Schwarzschild mass". 
Finally, the last term encodes model-dependent corrections to the universal term. 
At fixed time, asymptotic flatness of the metric functions constrains $e^{-\beta_0(L)}\leq 1$. This implies that, for $k_L>0$, the third term in \cref{MSmasscosmologygeneral} gives a subleading correction to $M(\eta)$ at any redshift $z = (1-a)/a$ and it decreases as $z$ grows.
It is important to stress again that  \cref{MSmasscosmologygeneral} is valid for any compact object (either singular or nonsingular) allowed by GR. In the following, we will apply separately our formula to the three possible cases: singular BHs, nonsingular BHs and regular horizonless (star-like) objects.

\subsection{Singular BHs}
For standard Schwarzschild BHs, the sum of the last two terms in \cref{MSmasscosmologygeneral} is identically zero and we are left with the purely cosmological contribution. This is true at every fixed instant of time, as it can be seen from \cref{alpha00static}, by substituting $e^{-\beta(r)} = 1- 2GM/r$. Thus, Schwarzschild BHs decouple from the large-scale cosmological dynamics. Recent investigations \cite{Cadoni:2023lqe} suggest that this is a quite general result, valid for generic, local and singular compact objects.
The coupling is, however, present for nonsingular objects. In the following, we will consider this latter possibility.

\subsection{Nonsingular BHs}
When the compact object is a regular BH, $L$ can be naturally identified with the radius of the apparent horizon (trapped surface). The subleading contribution in \cref{MSmasscosmologygeneral} will, thus, always be zero in these cases. This means that the cosmological mass coupling is universal for BHs and our formula \eqref{MSmasscosmologygeneral} becomes equivalent to \cref{Farraformula}, with $k=1$.

\subsection{Horizonless compact objects}
Let us now discuss the application of our mass formula to the  compact, horizonless solutions of a notable class of regular models with EoS $p_\varpar = -\rho$, which has been extensively used to construct regular GR solutions \cite{Cadoni:2022chn,Hayward:2005gi,Dymnikova:2004qg,bardeen,Beltracchi:2018ait,Nicolini:2005vd, Simpson:2019mud}. The latter are also of particular interest because, owing to their EoS, they are the GR solutions most similar to dark energy.

Using this EoS into the system \eqref{SystemEqFirstRegime}, we have ${\alpha} =-\beta_0$. In the following, we will focus on two paradigmatic cases: models endowed with a de Sitter core \cite{Cadoni:2022chn} and those with an asymptotically Minkowski core \cite{Simpson:2019mud}. All the latter are characterized by their ADM mass $\mathcal{M}$ and by an external length scale $\ell$, responsible for the smearing of the classical singularity. Depending on its value w.r.t. the Schwarzschild radius, these solutions can have either two or no horizons or a single degenerate one, representing the extremal configuration (for more details, see Refs.~\cite{Cadoni:2022chn, Simpson:2019mud}). The metric describing all these models can be written in the general form \cite{Cadoni:2022chn}
\begin{equation}
    e^{\alpha(r)} =1-\frac{2Gm(r)}{r} = 1-\frac{2G\mathcal{M}}{\ell} F\left(\frac{r}{\ell} \right)\, ,
    \label{generalmetricfunction}
\end{equation}
where the function $F$ encodes the information about the MS mass $m(r)$ at the fixed conformal time $\eta_i$-the conformal time at which the object forms couples to the cosmological background-and, thus, realizes the interpolation between the behavior near the core (regularizing the $r \sim 0$ singularity) and the asymptotic Schwarzschild solution at $r \to \infty$. 

With these few ingredients, we can now consider the horizonless configurations of the models \eqref{generalmetricfunction} and describe the behavior of the subleading, model-dependent term in \cref{MSmasscosmologygeneral}, $\mu(a) \equiv e^{-\beta_0(L)} a^{k_L} = e^{\alpha(L)}  a^{k_L}$, in all generality.

Since we only consider here the case of the absence of horizons, $e^{\alpha(L)}$ is always positive. Additionally, given the  asymptotic behavior at $r \sim 0$ and $r \to \infty$, it is always bounded by $1$ from above, whatever the value of $L$. Evaluating the latter is anyway essential to assess the sign of $k_L$, which in turn determines the variation of $\mu$ with the redshift. Since the density profile of the models considered here goes to zero only at $r\to \infty$, we cannot define a hard surface for these stellar objects. We can, however, define an effective radius, containing $\sim 99 \, \%$ of the ADM mass, namely $m(L)/\mathcal{M} = 0.99$. The consequence of this prescription is that $L$ is sufficiently big to ensure that \cref{generalmetricfunction}, evaluated at $r = L$, is well approximated by the Schwarzschild metric. This implies that $F'(L/\ell) < 0$. $k_L$ is, thus, positive for all models. Therefore, $\mu(z)$ starts at a maximum value, less than $1$, at $z = 0$ and then decreases with the redshift.

We confirmed these conclusions considering three specific models: a particular case of the Fan $\&$ Wang metric \cite{Fan:2016hvf} (recently revisited in \cite{Cadoni:2023nrm,Cadoni:2022vsn}), the Gaussian core regular model \cite{Nicolini:2005vd} and the asymptotically Minkowski core model of Ref.~\cite{Simpson:2019mud}. We noted that, for all these, the subleading correction $\mu(z)$ has an order of magnitude which is comparable with $1$ at $z=0$. However, while the magnitude of $\mu$ stays close to $1$ in a wide interval of redshifts for the first and last models (due to the slow fall of their density profiles at infinity), it decreases much faster for the Gaussian core model. 

\subsection{Can nonsingular compact objects contribute to the dark energy?}
If the redshift of BH masses is described by \cref{Farraformula} with $k=3$, as the observational evidence with elliptical galaxies  found in Ref.~\cite{Farrah:2023opk} seems to suggest, then BHs should give a contribution to the cosmological constant in the Friedmann equations (see Ref.~\cite{Farrah:2023opk}). On the other hand, if $k=1$, as we have shown to be the case adopting a GR framework to describe them, BHs contribute to the cosmological energy density as a curvature term, which redshifts as $\rho\sim a^{-2}$. Therefore, the interpretation of vacuum energy in terms of GR BHs is quite problematic, and would imply a nontrivial averaging of inhomogeneities on large, cosmological, scales (see Ref.~\cite{Cadoni:2020jxe}).   

\begin{figure}[t]
\centering
\includegraphics[width=\textwidth]{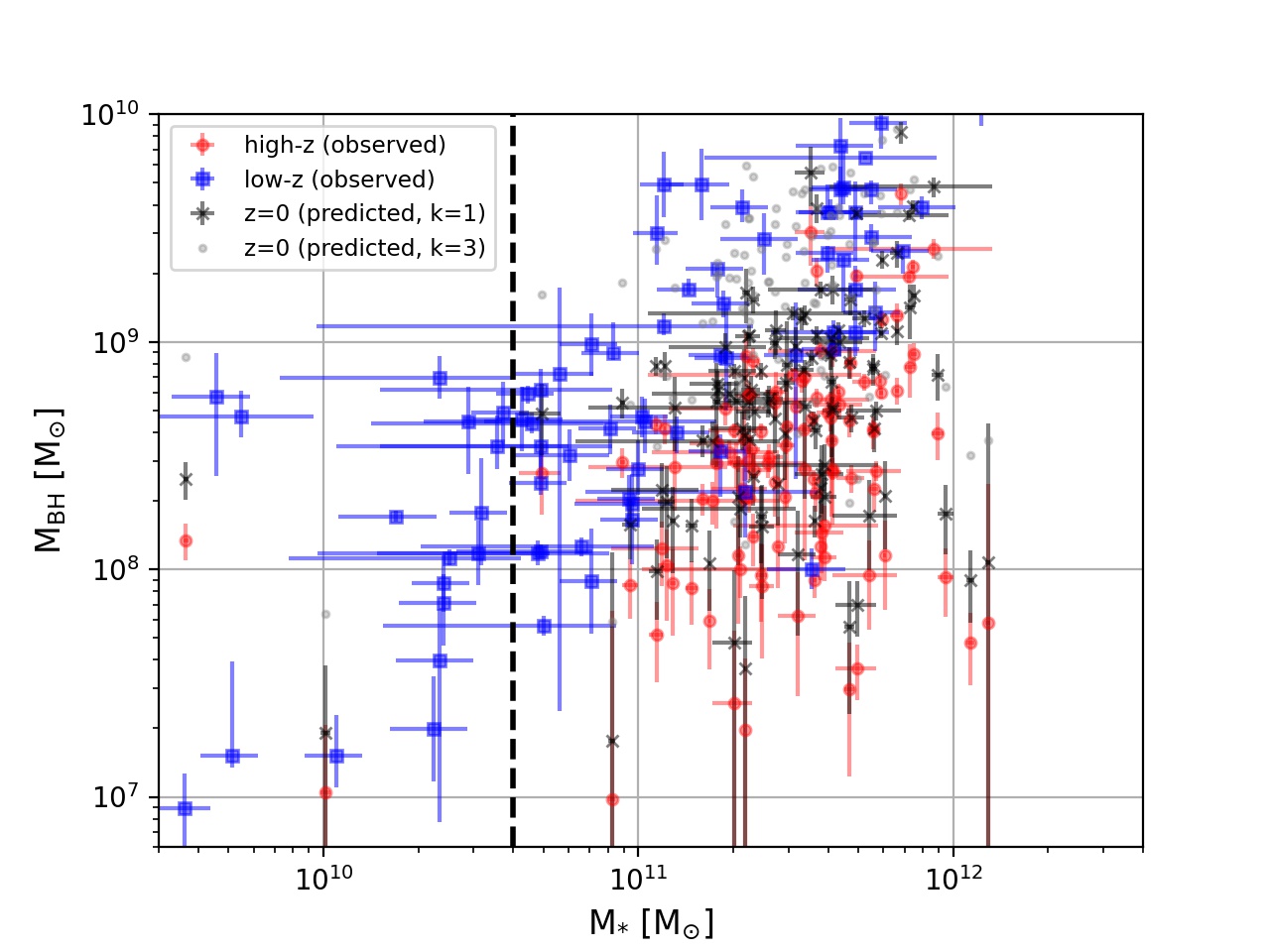}
\caption{Here we show the unshifted final high-redshift sample (\textit{i.e.}, assuming zero cosmological couplings) (red circles), as well as the cases when it is shifted with $k=1$ (black crosses), and $k=3$ (gray dots). 
In addition, we show the unshifted low-redshift sample from Table 2 of Ref.~\cite{Farrah_2023} (blue squares). The vertical dashed line represents the low stellar mass cut of $M_{*} > 4 \cdot 10^{10} \, M_{\odot}$ adopted in our data analysis.}
\label{fig:FinalPlot}
\end{figure}

\section{Constraints from the growth of supermassive BHs in elliptical galaxies} 

\begin{figure}[t]
\centering
\includegraphics[width=\textwidth]{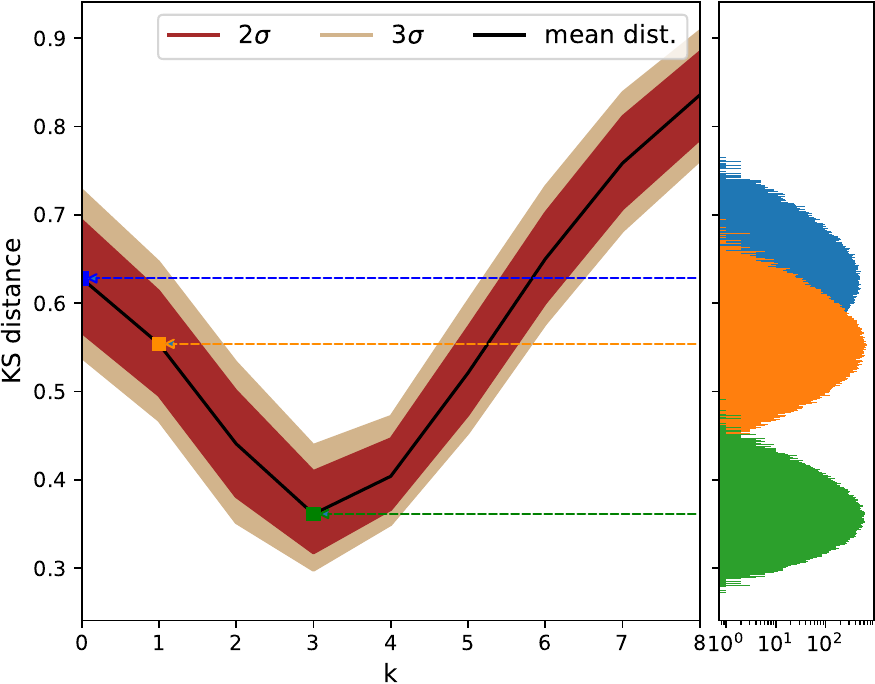}
\caption{The $95\, \%$ (dark red) and $99.7\, \%$ (brown) containment of the KS distance estimates for different $k$ values using $10^5$ realizations of low-redshift and high-redshift samples shifted to $z=0$. The right vertical panel shows the distribution of KS distances simulated using the uncertainty on the mass of the black hole ($\Delta \rm M_{\rm BH}$) and the uncertainty on the mass of the star ($\Delta \rm M_{*}$).}
\label{fig:Probdist}
\end{figure}

Here, we compare our theoretical prediction with the mass measurements of SMBHs in elliptical galaxies at different redshifts. As done in Ref.~\cite{Farrah_2023}, we select “red-sequence” elliptical galaxy hosts where the SMBH growth via accretion is expected to be negligible, and thus the measured BH mass growth with z can be considered exclusively due to cosmological coupling. We consider a set of low- ($0.0016 \leq z \leq 0.19$) and high-redshift ($0.8 \leq z \leq 0.9$) SMBHs. The low-redshift sample is directly taken from Table 2 of Ref.~\cite{Farrah_2023}. The high-redshift sample is built by cross-matching two AGN catalogues, one from the Wide-field Infrared Survey Explorer (WISE) survey~\cite{Barrows:2021riy}, and one from the Sloan Digital Sky Survey (SDSS) survey~\cite{rakshit2020spectral}. The relevant properties of the host galaxies,
such as, the star-forming rate (SFR), AGN reddening ($E_{b-v}$) and stellar mass ($M_\ast$) are taken from Ref.~\cite{Barrows:2021riy}, while the BH masses from Ref.~\cite{rakshit2020spectral}. 
The final high redshift sample ${M_{*}- M_{\rm{BH}}}$ used for our analysis is constructed following the procedure described below:

\begin{enumerate}
\item The objects from the above-mentioned catalogues are selected cross-matching their relative position (R.A. and Dec.) within $2^{''}$. The resulting sample consists of $160,005$ objects. 

\item Only objects within the redshift range $0.8 \leq z \leq 0.9$ are selected. This ensures that the measurements of the SMBH masses are estimated using the Mg II line. This cut reduces the sample to $605$ objects.

\item All objects for which $M_{*}$ and/or $M_{\rm BH}$ are undefined are removed, reducing the sample to $554$ objects.

\item A cut on the AGN reddening parameter $E_{b-v} < 0.2$ is applied. This ensures to avoid reddened AGN and select unobscured AGN. This cut reduces the sample to $510$ objects.

\item A selection on the SFR below a factor $5$ from the best-fit SFR-$M_{*}$ is adopted to ensure the quiescence of the selected galaxies. This is done via the relation obtained in Ref.~\cite{Speagle:2014loa} (Equation~(28)). This cut further reduces the sample size to $211$ objects.

\item Finally, in order to select only elliptical galaxies, we reject objects for which the bolometric luminosity of the spiral component is brighter than 5\% of that of the elliptical component\footnote{We make use of the following parameters from Ref.~\cite{Barrows:2021riy} for this cut: \textbf{L\_E\_best} (bolometric luminosity of the elliptical component, coloumn 24) and  \textbf{L\_Sbc\_best} (bolometric luminosity of the elliptical component, coloumn 27).}.
\end{enumerate}

The final high-$z$ $\{M_{*}- M_{\rm BH}\}$-sample used in our analysis consists of $108$ objects. It is shifted to $z=0$ using Equation~\eqref{Farraformula}, both with $k=1$ and $k=3$. Figure~\ref{fig:FinalPlot} shows the unshifted final high-redshift sample,~\textit{i.e.},~assuming zero cosmological coupling (red circles), as well as the cases when it is shifted with $k=1$ (Sample-A, black crosses), and $k=3$ (Sample-B, grey dots). 
In addition, we show the low-redshift sample from Table 2 of Ref.~\cite{Farrah_2023} with blue squares (Sample-C). 
Note that, in order to compare the high- and low-redshift data, also Sample-C requires to be shifted to $z=0$ via Equation~\eqref{Farraformula} with $k=1$ and $k=3$. However, being all objects included in Sample-C at a redshift close to zero, the resulting shift in the ${M_{\rm BH}}$-axis of Figure~\ref{fig:FinalPlot} is not visible by eye.

We now compute the probability that the two samples (Sample-A and -B) are drawn from the same sample as Sample-C using a two-dimensional Kolmogorov-Smirnov (KS) test~\cite{10.1093/mnras/202.3.615,1987MNRAS.225..155F}. 
In order not to bias the fit, we impose a low stellar mass cut of $M_{*} > 4 \cdot 10^{10} \, M_{\odot}$, due to the fact that in the high-redshift sample there are very few objects below this value as in~\cite{Farrah_2023}.

For a given value of $k$, we base our statistical analysis on the estimation of the minimum KS distance between the resulting low- and high-redshift samples, each of them shifted to $z=0$ with the given $k$-value. 

Per each $k$ spanning the interval $[0-8]$, we simulate $10^5$ random realizations assuming a $2$D Gaussian distribution for each point defined by $\left[M_{*}\pm\Delta {M}_{*}, M_{\rm BH}\pm\Delta {M}_{\rm BH}\right]$ in the $2$D plane shown in Figure~\ref{fig:FinalPlot}\footnote{$M_{*}$ and $M_{\rm BH}$ are the mean values of the distributions, while $\Delta { M}_{*}$ and $\Delta { M}_{\rm BH}$ the corresponding standard deviations.}. This ensures the inclusion of the uncertainty on the mass measurements into the analysis.
Per each $k$-value that we considered, we are thus left with a distribution of $10^5$ KS-distances, as shown in the right panel of Figure~\ref{fig:Probdist} for the $k=0$, $1$, $3$ cases.
Given a value of $k$, we derive the mean distance, $2\sigma$ ($95\%$), and $3\sigma$ ($99.7\%$) confidence limits. By iterating this procedure in the range $k = [0,8]$ we obtain the allowed $2$- (dark red patch) and $3$-$\sigma$ (brown patch) bands as a function of $k$, as reported in the left panel of Figure~\ref{fig:Probdist}.
Figure~\ref{fig:Probdist} shows that the preferred value of $k$ is around $3$, since it corresponds to the smallest KS-distance. On the contrary, both the cases $k=0$ and $k=1$, consistent within $2\sigma$ with each other, are in tension with the obtained best-fit value of $k$.

While the preference for $k \simeq 3$ is consistent with the results of the analysis in Ref.~\cite{Farrah:2023opk} with a similar data set from WISE/SDSS, our theoretical prediction is consistent with the BH mass-coupling test performed using higher redshift AGN and quasar data from the JWST \cite{Lei:2023mke} survey, the result of which is quite in tension with that of Ref.~\cite{Farrah:2023opk}. The posterior probability distribution for $k$ obtained in Ref.~\cite{Lei:2023mke} is centered around a best-fit value $k=-0.94\pm 1.19$ at $68\%$ confidence level, deviating from $k=3$ at more than $3\sigma$.
Our theoretical prediction, $k=1$, is compatible within $1 \sigma$ with the JWST data. 
Let us note that the results presented in Ref.~\cite{Lei:2023mke} are obtained with three AGN in early-type host galaxies, namely, CEERS01244, CEERS00397, and CEERS00717, and the conclusions are mainly driven considering the object CEERS01244 at redshift $z=4.8$. However, the JWST is promisingly expected to increase the sample of a such measurements in the upcoming years.

The approach adopted in this work relies on the comparison between data sets of BH/early-type galaxies from different redshift regimes. Systematic uncertainties, possibly arising from the different observations and observables used to estimate the mass of the BH and the stellar mass of the host galaxy, may not be completely under control. For example, the JWST results come from a sample of stellar masses smaller than $10^{10}$ $M_\odot$. This range of masses is excluded from the analysis of Ref. \cite{Farrah:2023opk}, making our comparison between the two data sets more complicated.

\section{Final remarks}
In this letter, we have shown that GR, in its most general settings allowing nonsingular BH solutions, predicts a universal cosmological redshift of their masses, determined by the local curvature term. This consequently implies a linear scaling of the mass with the scale factor $a$, \textit{i.e.}, $M \propto a$. In the case of horizonless regular compact objects, we have found a subleading, model-dependent, correction to the universal curvature term.    

In order to check our theoretical prediction, we have performed a statistical analysis (KS-distance test) of the data of the mass growth of SMBHs in elliptical galaxies. We have compared the observed sample at low redshift with a sample at high-redshift, for several values of $k$ ranging from $0$ to $8$. We have found the smallest KS-distance for $k=3$, in accordance with the results of Ref.~ \cite{Farrah:2023opk}. On the other hand, our theoretical prediction, $k=1$, is compatible within $1 \sigma$ with the results of the statistical analysis performed in Ref.~\cite{Lei:2023mke}, based instead on the data of high-redshift AGNs detected by the JWST. 
Small values of $k$ are also confirmed by the recent findings obtained with LIGO/Virgo/KAGRA data of binary BH mergers, which indicate a mild preference towards $k\sim 0.5$~\cite{Croker:2021duf}. 

In view of this rather intricate situation, we need to consider other astrophysical observations and data, both of SMBHs and binary mergers, in order to assess which value of $k$ is favoured by these samples.
Our work will largely benefit from future observations. As underlined in Ref.~\cite{Lei:2023mke} the JWST will observe a large number of high redshift galaxies and AGNs identifying a larger number of early-type host galaxies with higher stellar mass in the range of masses of the catalogs used in the present analysis. The improved sensitivity of GW observatories, such as LIGO/Virgo/KAGRA, will enable us to directly measure the masses of an increasingly large sample  of stellar-mass BHs. Even more strikingly, the next generation of GW detectors, such as the Einstein Telescope and Cosmic Explorer, will enormously increase the number of detections of stellar mass binary BHs ($10^5$) per year up to higher and higher redshifts. Among these detections, a large fraction will have accurate mass estimates (see Ref.~\cite{Branchesi2023} for details). Moreover, the Laser Interferometer Space Antenna (LISA) will enable the access to massive BH coalescences. More precise mass estimates, broader coverage of mass range, and the use of different messengers will be crucial to constrain the cosmological coupling strength while keeping under control observational biases and possible systematics.

A preference for $k = 0$ from future observations will imply that nonsingular GR BHs are hardly compatible with the observational evidence. If data will, instead, favour $k=1$, \textit{i.e.}, a non-zero cosmological coupling, apart from being important \emph{per s\`{e}}, this result could represent striking evidence that GR BHs are indeed nonsingular objects. If this will be the case, we will still have to discriminate between objects endowed with a horizon from horizonless (stellar-like) ones. To this end, we will need a higher level of precision, sufficient to detect the subleading corrections derived in this work. Conversely, a confirmation from the data of either the $k=3$ result of Ref.~\cite{Farrah:2023opk} or the $k=-1$ result of Ref.~\cite{Lei:2023mke} will, instead, imply that BHs are nonsingular objects, which however do not admit an effective description in terms of solutions of Einstein's field equations. This will give a strong indication of the existence of new gravitational physics beyond GR at astrophysical scales.

\section*{Acknowledgements and data availability}
We are thankful to Elias Kammoun, Eleonora Loffredo, Andrea Maselli, and Gor Oganesyan for useful discussions. We thank Scott Barrows for providing the AGN catalog presented in Ref.~\cite{Barrows:2021riy}.\\
We acknowledge the use of {\tt Matplotlib}~\cite{Hunter:2007}, {\tt NumPy}~\cite{harris2020array}, {\tt SciPy}~\cite{Virtanen:2019joe}, 
{\tt{asymmetric\textunderscore uncertainty}}\footnote{\url{https://github.com/cgobat/asymmetric_uncertainty}}~\cite{2022ascl.soft08005G}, {\tt{AstroPy}}~\cite{astropy:2022}, {\tt ndtest}\footnote{\url{https://github.com/syrte/ndtest}}, {\tt 2DKS}\footnote{\url{https://github.com/Gabinou/2DKS}}. The results of our data analysis are fully reproducible by making use of publicly available data catalogs\footnote{\url{https://archive.stsci.edu/hlsp/candels/cosmos-catalogs}}~\cite{rakshit2020spectral}. BB and MB acknowledge financial support from the METE project funded by MUR (PRIN 2020 grant 2020KB33TP). 

\bibliography{refs}
\end{document}